\documentclass{sig-alternate-05-2015}

\usepackage{color}
\usepackage{graphicx}
\usepackage{textcomp}

\usepackage{array}
\usepackage{xspace}
\usepackage{hyperref}
\providecommand{\tabularnewline}{\\}
\newcommand{\zebra}{$\mathit{Zebra-RFO}$\xspace}

\makeatletter
\def\@copyrightspace{\relax}
\makeatother

\newif\ifeurasia
\eurasiafalse
\ifeurasia
\fi

\begin{document}

% Copyright
%\setcopyright{acmcopyright}
%\setcopyright{acmlicensed}
%\setcopyright{rightsretained}
%\setcopyright{usgov}
%\setcopyright{usgovmixed}
%\setcopyright{cagov}
%\setcopyright{cagovmixed}

%\CopyrightYear{2016} 
%\setcopyright{othergov}
%\conferenceinfo{GAIA,}{August 22-26, 2016, Florianopolis, Brazil}
%\isbn{978-1-4503-4423-4/16/08}\acmPrice{\$15.00}
%\doi{http://dx.doi.org/10.1145/2940157.2940160}

%
% The code below should be generated by the tool at
% http://dl.acm.org/ccs.cfm
% Please copy and paste the code instead of the example below. 
%

\bibliographystyle{plain}
\title{Open and Regionalised Spectrum Repositories for Emerging Countries}
\numberofauthors{6} %
\author{
% You can go ahead and credit any number of authors here,
% e.g. one 'row of three' or two rows (consisting of one row of three
% and a second row of one, two or three).
%
% The command \alignauthor (no curly braces needed) should
% precede each author name, affiliation/snail-mail address and
% e-mail address. Additionally, tag each line of
% affiliation/address with \affaddr, and tag the
% e-mail address with \email.
%
% 1st. author
\alignauthor Andr\'es Arcia-Moret\\
       \affaddr{University of Cambridge, UK}\\
       \email{\normalsize andres.arcia@cl.cam.ac.uk}
% 2nd. author
\alignauthor Arjuna Sathiaseelan\\
       \affaddr{University of Cambridge, UK}\\
       \email{\normalsize arjuna.sathiaseelan@cl.cam.ac.uk}
% 3rd. author
\alignauthor Marco Zennaro\\
       \affaddr{ICTP, Italy}\\
       \email{\normalsize mzennaro@ictp.it}
\and  % use '\and' if you need 'another row' of author names
% 4th. author
\alignauthor Freddy Rond\'on\\
       \affaddr{University of Los Andes, Venezuela}\\
       \email{\normalsize frondon@ula.ve}
% 5th. author
\alignauthor Ermanno Pietrosemoli\\
       \affaddr{ICTP, Italy}\\
       \email{\normalsize ermanno@ictp.it}
% 6th. author
\alignauthor David Johnson\\
       \affaddr{CSIR, South Africa}\\
       \email{\normalsize djohnson@csir.co.za}
}

\maketitle

\begin{abstract} 

% this is the abstract from David's email:
%TV Whiste Spaces have recently been proposed as an alternative to alleviate the spectrum crunch, characterised by the need to reallocate frequency bands to accommodate the ever-growing demand for wireless communications. TV broadcasting spectrum allocation was established many decades ago when TV was the most popular wireless device. Also, spectrum efficiency was very poor due to the need for multi-frequency broadcast networks and adjacent channel guard bands to avoid co-interference. Furthermore, TV spectrum allocations were typically made nationwide, without regard to the differences in the population density between urban and rural areas. The launch of Digital Terrestrial TV has resulted in improved spectrum efficiency opening up more free TV channels, especially in rural areas. As a consequence, there are substantial portions of spectrum available for alternative usage. To leverage this available spectrum or TVWS, it is paramount to perform measurements in many places and publish this information as widely as possible. In this paper, we discuss the motivations and challenges for collecting spectrum measurements in developing regions and discuss a scalable system for the crowds to gather and provide access to WS information through open and regionalised repositories.

TV White Spaces have recently been proposed as an alternative to alleviate the spectrum crunch, characterised by the need to reallocate frequency bands to accommodate the ever-growing demand for wireless communications. In this paper, we discuss the motivations and challenges for collecting spectrum measurements in developing regions and discuss a scalable system for communities to gather and provide access to White Spaces information through open and regionalised repositories. We further discuss two relevant aspects. First, we propose a cooperative mechanism for sensing spectrum availability using a detector approach. Second, we propose a strategy (and an architecture) on the database side to implement spectrum governance. Other aspects of the work include discussion of an extensive measurement campaign showing a number of white spaces in developing regions, an overview of our experience on low-cost spectrum analysers, and the architecture of \zebra, an application for processing crowd-sourced spectrum data.

\end{abstract}


   %Añadir este comando para que inserte lo anterior.

\section{Introduction}
\label{intro}

The convenience of wireless networks in supporting mobility and ease of deployment has made them extremely popular. These networks convey data in the order of a couple of exabytes per month, and in the next five years, this number is expected to grow at least one order of magnitude. A natural consequence of this tendency is a congested wireless spectrum in the band for cellular communications as well as the licence free ISM band, thus creating the so-called spectrum crunch. 

To keep track of primary users and incumbents use of the wireless spectrum, regulators in emerging countries use manual and static databases. However, there are intermittent legal users (e.g., UHF microphones), unaccounted legal users, and rogue users that utilize the spectrum with no control and can act as potential interferers \cite{bahl}. This circumstance is a clear opportunity for regulators and local authorities to promote regionalized (i.e., distributed) repositories for keeping track of used and unused frequencies, boosting efficient use of wireless spectrum. One of the clear applications involves managing white spaces inside buildings as spectrum availability varies from building to building \cite{ying}. This allows fine-grained management of spectrum and improve spectrum efficiency. Moreover, this is a promising way of not only tackling the spectrum crunch, i.e., through an appropriate assessment of spectrum usage, but this also allows experimentation with long-distance point-to-point links in TV WhiteSpace (TVWS). Applications that make use of this approach are long-distance backhaul links, emergency communication links, Public Protection and Disaster Relief provisioning \cite{holland}.

Regional repositories will allow people and governments to cooperate, paving the way to alternative wireless network deployments bringing Internet connectivity, especially in emerging regions. Successful examples of such networks operating in the free spectrum are: GuifiNet\footnote{\footnotesize https://guifi.net/}, entirely built by independent organizations or, long distance TV White Spaces deployments in the UHF band in Africa\footnote{\footnotesize http://www.carlsonwireless.com/white-space-hotspot/}. Based on these success stories, along with an appropriate use of the spectrum, interested parties should also incentivize the creation of community wireless service providers, better placed to understand the local people's needs \cite{sathiaseelan}. Such networks enable better content delivery and adequate support for a local production of content and services. 

Recent wireless technologies such as TVWS can be deployed if there is enough information about unoccupied portions of the spectrum. TVWS networks can be deployed in rural and remote areas more clearly because of the large amount of available TV spectrum and because they are well suited to long distances and provide a cost-effective solution \cite{mikeka}. 

%However, a successful deployment depends on the availability of spectrum, but measuring these dynamics has always been an expensive task \cite{arcia}. The cost of spectrum analyzers is in the order of thousand of dollars and the processing of information generated by these devices is not oriented towards understanding the available frequencies of interest (or white spaces) within a geographical region. 

To understand the current occupation of the UHF and ISM band, we count on open low-cost systems for capturing and processing spectrum dynamics in extensive areas\footnote{\footnotesize Zebra-RFO has been developed by Andr\'es Arcia-Moret and Freddy Rond\'on at the University of Los Andes.}. We incentivize people to be aware of the local occupation of their spectrum so they can be in charge and vigilant with their spectrum resources. To meet this objective, we discuss \textit{www.zebra-rfo.org}: a web system with collaboration capabilities similar to social networks, able to organize long measurement campaigns to visualize the occupation of the spectrum. \zebra also offers the possibility of editing measurement campaigns to isolate different areas of interest (i.e., rural, urban, suburban), and also conveniently represent the rough occupation of large portions of the spectrum in UHF band and ISM band, both of high interest for bringing the next billion people on-line.

% Here one could argue that there are many studies that show that the WiFi  approach of shared spectrum has been much more efficient in terms of spectrum utilization than dedicated allocation like in cellular.

\textbf{Dealing with regulations.} Although the superior propagation characteristics of sub 1-GHz frequencies make them good candidates for alleviating the spectrum crunch, one of the major obstacles to making use of these frequencies for Internet connectivity is to persuade regulators of the benefits for rural populations - especially those in developing countries. In this context, there is a need for low-cost spectrum monitoring in TVWS. 

More recently, the ITU has recognized that TV broadcasting spectrum is currently being underutilised, and has recommended reallocating the frequencies above 694 MHz from TV broadcasting to cellular service. Nevertheless, the frequencies from 470 to 694 MHz offer plenty of opportunity for TVWS in rural areas, where the broadcaster can make a business case for just a few channels leaving the rest of the spectrum empty. Whereas in urban areas, there are less empty spaces in the TV band, but cellular operators can afford to deploy many base stations to meet the higher consumer demand. Interestingly, offloading wired broadband networks onto cellular networks has been demonstrated as an option for enhancing the performance of particular applications \cite{rossi}. 

Successful trials in Malawi \cite{mikeka} suggest that collecting spectrum dynamics with low-cost equipment can help in convincing regulators of the sub-utilisation of the spectrum. Moreover, monitoring costs are cut dramatically, and a first rough view of the spectrum occupancy is arguably better than the static (manual) approach \cite{brown}.

\textbf{Open and Regionalised Repositories.} It is in the interest of the government (through regulatory authorities) to manage and control the spectrum in populated regions. Since spectrum is a high-value resource and each country has sovereign control over their spectrum apart from meeting ITU guidelines on inter-border interference (e.g. GE-06 conditions for use of broadcasting bands), spectrum governance structures are always well defined within each country. However, in developing regions spectrum governance is often challenging due to: (1) inefficient management of the spectrum occupancy -- having stale information maintained by semi-automatized databases, (2) governmental unwillingness to be compliant to international structures and regulations, such is the case of the under-compliance to the ITU recommendations for spectrum occupancy in Latin America (see \url{http://goo.gl/Znmjna}), (3) Very long approval times to license portion of the spectrum, even when those portions are well known to be unoccupied. (4) A need to improve spectrum occupancy data-access speed.

%\textbf{Challenges.} A number of challenges have to be overcome when designing a regionalised repository managing scalable data and different specific regulations imposed by governments: massification of the spectrum repositories, with the aid of low-cost spectrum monitoring devices; overcoming governmental restrictions, i.e., attending different rules for assigning spectrum and allowing access to spectrum occupation data; departure from ITU recommendation for spectrum allocation in developing regions, most of the time giving no strong reasons, e.g., 20\% at best in Latin America for ITU recommendations on spectrum usage\footnote{\footnotesize http://goo.gl/Znmjna}; accessibility and understanding of occupation information by lay people so that spectrum information is not of exclusive use of (business-oriented) primary users; trade-off between the "timely" capture versus "mobile" capture of the spectrum. Resolution of the data can impose radical scale restrictions when fine grained monitoring capabilities are needed. paper
% \cite{3-4 SECs to access spectrum}

The above-stated situation led us to propose and discuss in this paper the use of open and regionalised spectrum repositories. These repositories are intended to provide a common place for people and governments to understand the occupancy of the spectrum and to meet every party's capabilities and restrictions. We address a fundamental trade-off: on the one hand, we assume that crowds are interested in measuring spectrum occupancy to deploy self-sustained community networks\footnote{\footnotesize https://datatracker.ietf.org/doc/draft-irtf-gaia-alternative-network-deployments/} and, on the other hand we suppose that governments have enough resources to process and govern the crowd-sourced data and to get it back to communities through a common \textit{official} interface. 

%\xxx{include possible uses of the data: governmental perspective of controlling the spectrum, the potential  use for forecasting occupation with machine learning \cite{vinicius}, use case of empowering people with information}

\section{State of the Art}

There has been a recent interest in building central repositories for observing white spaces across the globe. Microsoft Spectrum Observatory\footnote{\footnotesize https://observatory.microsoftspectrum.com} and Google Spectrum Database\footnote{\footnotesize https://www.google.com/get/spectrumdatabase/}, happen to have their own initiative and interface to massively collect spectrum dynamics. Different from our case, these early projects have been designed for the developed world and with privative design premises, such as a limited number of queries on the available data or centralised control of the collected data. We are contextualising open and regionalised repositories as a collaborative system between people and regulators. Furthermore, the whole collection workflow is being developed out of the experience on measuring white spaces in the developing world, thus our emphasis on low-cost equipment to measure and crowdsource, to eventually negotiate with regulators, in cases such as the deployment of alternative networks. 

%Other projects considering access to central repositories have reported query times in the order of 3 s \cite{majid}, considered inappropriate for mobile devices willing to profit from a dynamic strategy for accessing the spectrum. 

As explained in Section~\ref{sec:low-cost-collection} we have explored a handful of low-cost devices to promote the massive collection of the spectrum and at the same time to come up with a standard representation in a common repository.

There exist many architectures proposing the separation of the spectrum regulations into layers, especially for dynamic channel/frequency selection (DCS/DFS). These architectures are intended to protect the primary user in Unlicensed spectrum sharing cases. Irnich et al. \cite{irnich} proposes a spectrum sharing Toolbox (SST) to provide additional ways of authorizing spectrum usage - envisioning 5G scenario. The SST consists of three layers of operation for dealing with dynamic requests for the spectrum usage in a range of relevant spectrum sharing scenarios. A first toolbox layer containing among other components, a coordination protocol, a Geo Location Database, and a Spectrum Broker for fine-grained coordination of sharing. A second layer that provides the proper mechanism to ensure the lowest sharing overhead among primary users (peer operators) as well as unlicensed users, and finally, a third layer dealing with the regulatory framework (e.g., for assuring predictable QoS). In short, a layered approach is proposed for dealing with different use cases and distributed responsibilities of the spectrum sharing component. Hence, creating the niche for a scalable and distributed proposal for spectrum sharing.

There are pilot trials for assessing the network performance on White Spaces, like the one discussed by Holland et al. \cite{holland}. Authors propose communication with the Ofcom web listing of Geo-Location DBs, and communication with the Fairspectrum GLDB. This fact contrast with the notorious need for a layered and systematised approach as proposed in \cite{irnich} since the Ofcom approach is a centralised approach based on specific use-cases \cite{ofcom10}.

Other approaches addressed recently by the FCC corresponds to the low-powered network technologies like small cells in 5G \cite{chang}. The FCC encourages the design of a three-tiered spectrum management user-oriented system namely Spectrum Access System, similar to a TV Database. The top tier goes for the dynamic incumbents, the second tier for a Priority Access Licence users, and the third tier for a generalized authorized users. This architecture is intended to force secondary users to accurately back-off in the spatial regions of interest. Users in the 3rd tier have to protect themselves whereas users with Priority Access can safely have one or more allocated chunks. However, users belonging to the Citizen Broadband Service Devices class will require the system to manage the secondary user activity and minimize the interference to priority users and incumbent operations. Authors propose an architecture that separates a commercial system and a federal system to scale up by protecting the primary users and, at the same time, giving a service to secondary users.

Along with aforementioned architectural efforts, there also exists standardization efforts at the Internet Engineering Task Force (IETF) such as the RFC 7545 \cite{RFC7545} and 6953 \cite{RFC6953}. Based on the premise of government as the regulator, these documents aim at proposing common ground for white space access (management) and spectrum information retrieval. These guidelines are also intended to protect primary users that use the spectrum during limited time periods. They specify an agnostic interface in terms of transmission media, spectrum organization, and purpose, as well as flexible and extensible data structures modeling different use-cases, e.g., white space serving as backhaul or rapid deployments during emergencies. 

%\xxx{@Marco: include a refurbished version of the survey of measurements} This is the survey of Brown et al. \cite{brown}

\section{Low-Cost Spectrum Collection Process}
\label{sec:low-cost-collection}

Besides the reasons stated in Section~\ref{intro} concerning the organisational and potential social impact of a decentralised spectrum crowd-sourcing system, it is well known that spectrum analysers are expensive, difficult to transport, and they lack an appropriate interface to collect continuous spectrum activity (as suggested by the detector approach \cite{brown}). This situation motivated us to develop low-cost and long-term monitoring solutions. We have worked on assembling several such devices for the sub 1-GHz band: WhispPi monitor using a Raspberry Pi to interface RFExplorer \cite{arcia}, ASCII32 monitor using SI4313/Arduino \cite{zennaro}, and an Android interface for RFExplorer \cite{rainone}. These solutions have been worked out below the 400 US\$ price. However, they propose a heterogeneous vision of the spectrum usage due to the inherently different configurations, i.e., they use different antennas, various radio chips sets and sampling rates, in different geographical positions for capturing spectrum activity. Moreover, programmers tend to customize the capturing of the spectrum activity into different file formats, creating a need for a uniform format. In our proposal, defined in the next section, we define a simple JSON format - details can be seen at \url{http://www.zebra-rfo.org}. In this portal, several conversion scripts can also be used to interface the different aforementioned low-cost devices. 

\begin{figure}
\includegraphics[scale=0.21]{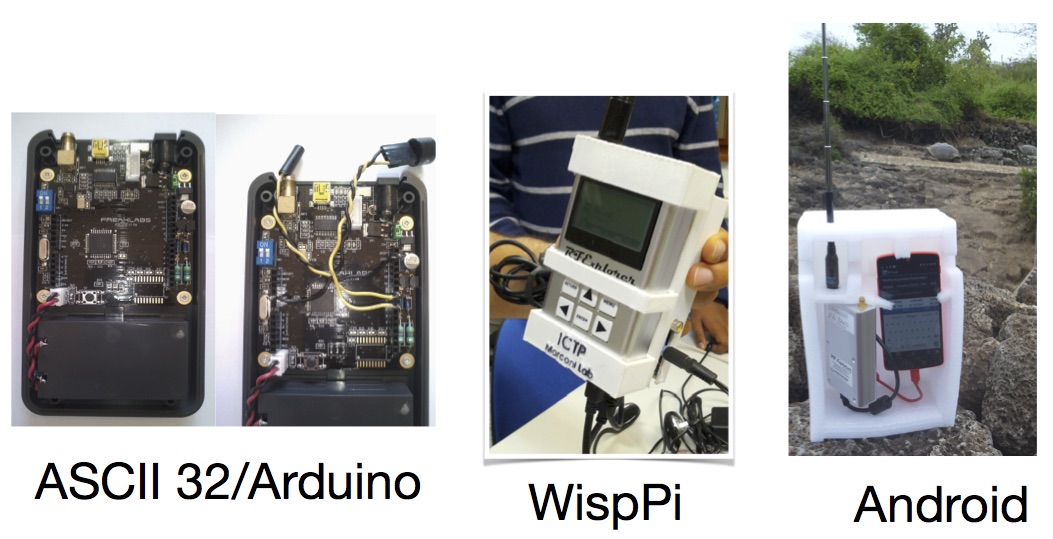}
\caption{Low-cost devices used to collect spectrum activity}
\label{fig:low-cost-devices}
\end{figure}

Fig.~\ref{fig:low-cost-devices} shows the different low-cost devices used. From left to right, we see the ASCII 32, WhisPi device, and the Android interface to the RFExplorer. They appear cost-ordered from cheapest to most expensive, and at the same time, the order follows their accuracy, from the least to the most accurate. From this group of devices, we have observed that for future designs, one should have better control on sampling rate, i.e., adapted base-rate and on-demand adaptive rate while obtaining mobile measurements. These demands are feasible characteristics when interfacing the RFExplorer through the Android OS, since many embedded sensors can help in such task (i.e., using the accelerometer or the GPS).

\section{A Regionalised Repository}

\begin{figure}
\includegraphics[scale=0.34]{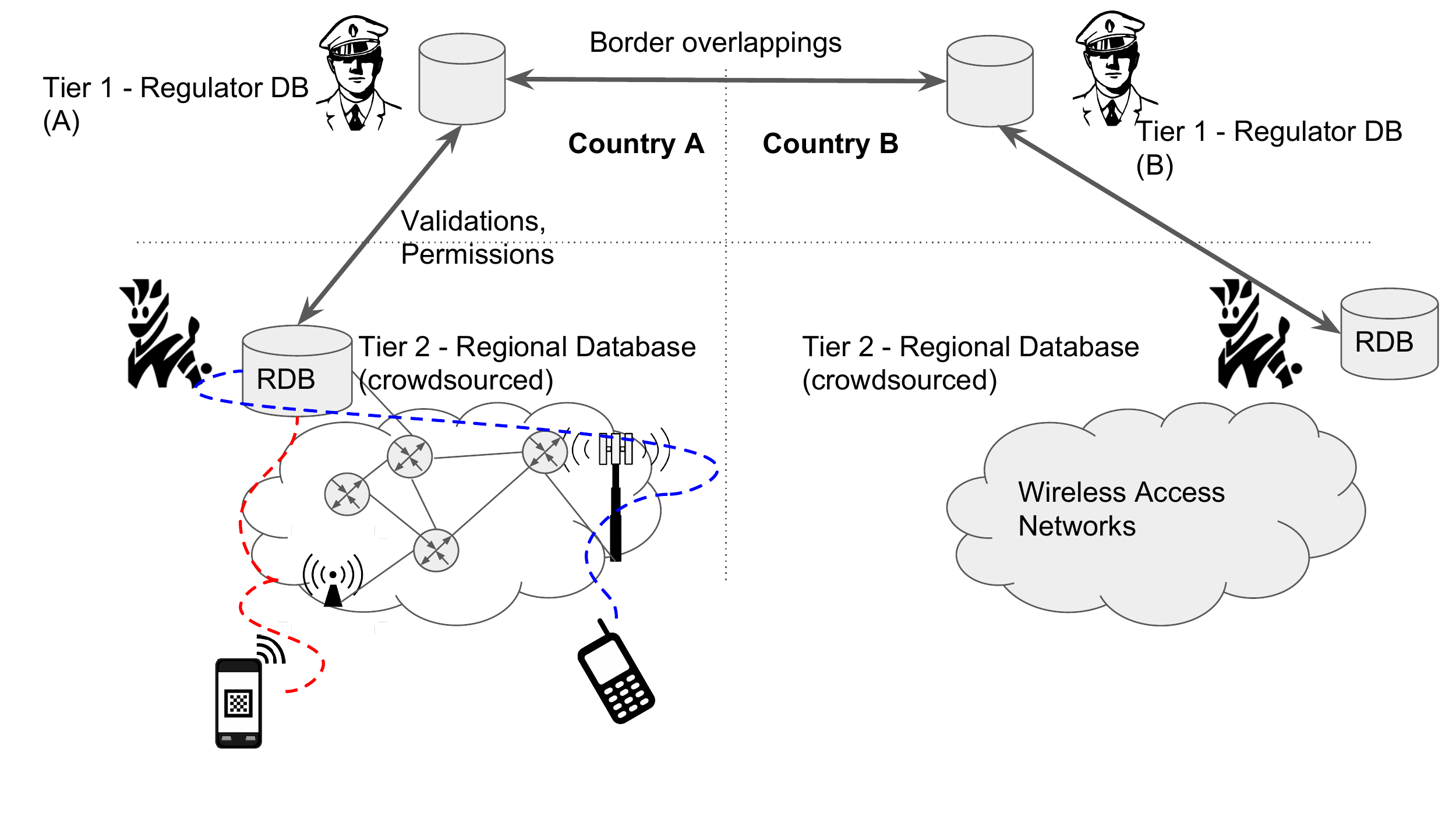}
\caption{Strawman architecture for Spectrum Governance}
\label{fig:arch-spec-gobernance}
\end{figure}

\begin{figure}
\includegraphics[scale=0.43]{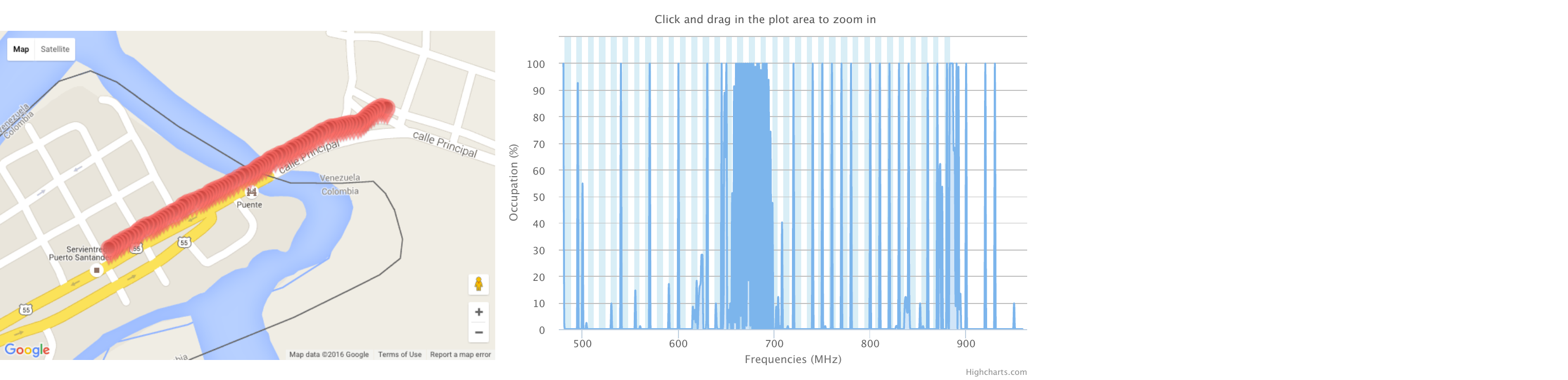}
\caption{Spectrum collection in Venezuela/Colombia border, accounting for a crowded semi-urban area}
\label{fig:colombia-venezuela}
\end{figure}

In this section, we present a prototype for a scalable and regionalised repository called \zebra: http://www.zebra-rfo.org, an open initiative to collect spectrum fingerprints and a social platform to incentivize a crowd-sharing approach for collection. The system provides data organization and visualization capabilities that allow later post-processing. \zebra offers capabilities such as a convenient edition of the geo-tagged journeys to get rid of potential biases introduced by the mobile collector speed (see Section~\ref{workflow} for further discussion). Moreover, editing the data allows one to isolate and categorize, with the aid of a visual interface, specific portions of the collected data. This capability allows filtering an area of interest, such as a well-defined urban area or, a rural area in which a TVWS network could be deployed (see Fig.~\ref{fig:colombia-venezuela}).

Fig.~\ref{fig:arch-spec-gobernance} shows a strawman architecture for our proposed spectrum governance system. This is a two-tiered architecture intended to separate crowd-sourced collections into regional databases and the authoritative feedback provided by the regulator in a different tier. There are several challenges for this architecture, namely, (1) the design of the inter-tier data exchange for validation purposes (reporting rogue users, as well as potential white spaces in specific sub-zones), (2) the definition (and design) of the protocols for intra-tier communication. This case corresponds to the inter-regulator communication for which we foresee cases of cross-border interference\footnote{\footnotesize The Geneva 2006 frequency plan (GE06) focused entirely on minimising cross-border interference of digital television. It covers Africa and Europe: http://goo.gl/21Y6Hu}. Such is the case of San Antonio, Venezuela, and Cucuta, Colombia urbanized border depicted Fig.~\ref{fig:colombia-venezuela} or, similarly in the rural Lilongwe, Malawi from where UHF Mozambique's signal could be perceived during our experience in the first TVWS deployment in Malawi \cite{mikeka}. In these cases, as suggested in Fig.~\ref{fig:colombia-venezuela} frequencies overlap, and operators have to agree on the virtual limits of their coverage. On the other hand, in the crowd-sourced spectrum tier, there can be overlapping spectrum usage that has to be dynamically solved, since mobile devices are likely to hand-off from one database to the next while moving.

\textbf{Regional Storage System.} As discussed in the previous section, our regional storage system poses protocol design challenges concerning the Inter-regulator Database synchronisation and Community-to-Regulator negotiations. However, we consider that a discussion in that direction is outside the scope of this paper, since we assume that the regulator has enough resources to provide backhaul connectivity, sufficient computational resources (i.e., in the cloud) and, legal, human expertise to process spectrum measurements. An example of how intensive computation for improved usage of the spectrum can be found in \cite{arciaWons}, we expose in detail the architectural role of an authoritative entity and an estimation of the required computational resources.

On the other hand, we discuss in this section a lower scale system. Our intention is to decrease the latency for accessing spectrum collections to empower communities with information on spectrum usage. This system has been tested on low-performance virtual machines on the cloud in two different providers, (a) an Amazon EC2 service providing the lowest ranked machine\footnote{\footnotesize \textbf{t1.micro:} 0.613 GB mem., low network performance, 1 vCPU and EBS disk service} under free tier and (b) the Cambridge Cloud Service with improved performance\footnote{\footnotesize \textbf{m1.medium:} 4 GB mem., high network performance, 2 vCPU and 40 GB of disk}. This selection has been made to emulate low-cost commodity equipment conditions. Our preliminary performance measurements report under a second time for the most expensive processing task for a 10 Km journey under (a) conditions and similar report for a 4x denser collection under (b) conditions.

%\begin{figure}
%\vspace{0.5cm}
%\includegraphics[scale=0.5]{img/system-architecture.pdf}
%\caption{\zebra system architecture}
%\label{fig:sys-arch}
%\end{figure}

%Fig.~\ref{fig:sys-arch} shows the \zebra system architecture. 

\textbf{System Architecture.} Similar to Chang et al. \cite{chang}, we use a system based on HTTP message exchange using JavaScript  Object Notation (JSON).  \zebra has been designed as a scalable and portable architecture based on services and RESTful interface so that we can deal with different clients (i.e., desktops as well as mobiles). The architecture is composed of 3 layers. The first layer, named presentation, is represented by the client. We use simple HTTP REQ/RESP messages to upload or download spectrum information. The second layer, the business, is represented by the server processing HTTP requests, who in turn will select a controller (i.e., uploader, filter, query, validator), depending on the URL and the HTTP request method. The so-called controller will implement the business logic included the communication with the third layer, named data layer which will store, in an adequate format, the spectrum data and eventually can reply to the client requests with JSON objects containing processed spectrum data.

%\xxx{missing to describe the synchronous and asyncronous processes in zebra}

\subsection{Spectrum Collection Workflow}
\label{workflow}

\begin{figure}
\vspace{0.5cm}
\includegraphics[scale=0.5]{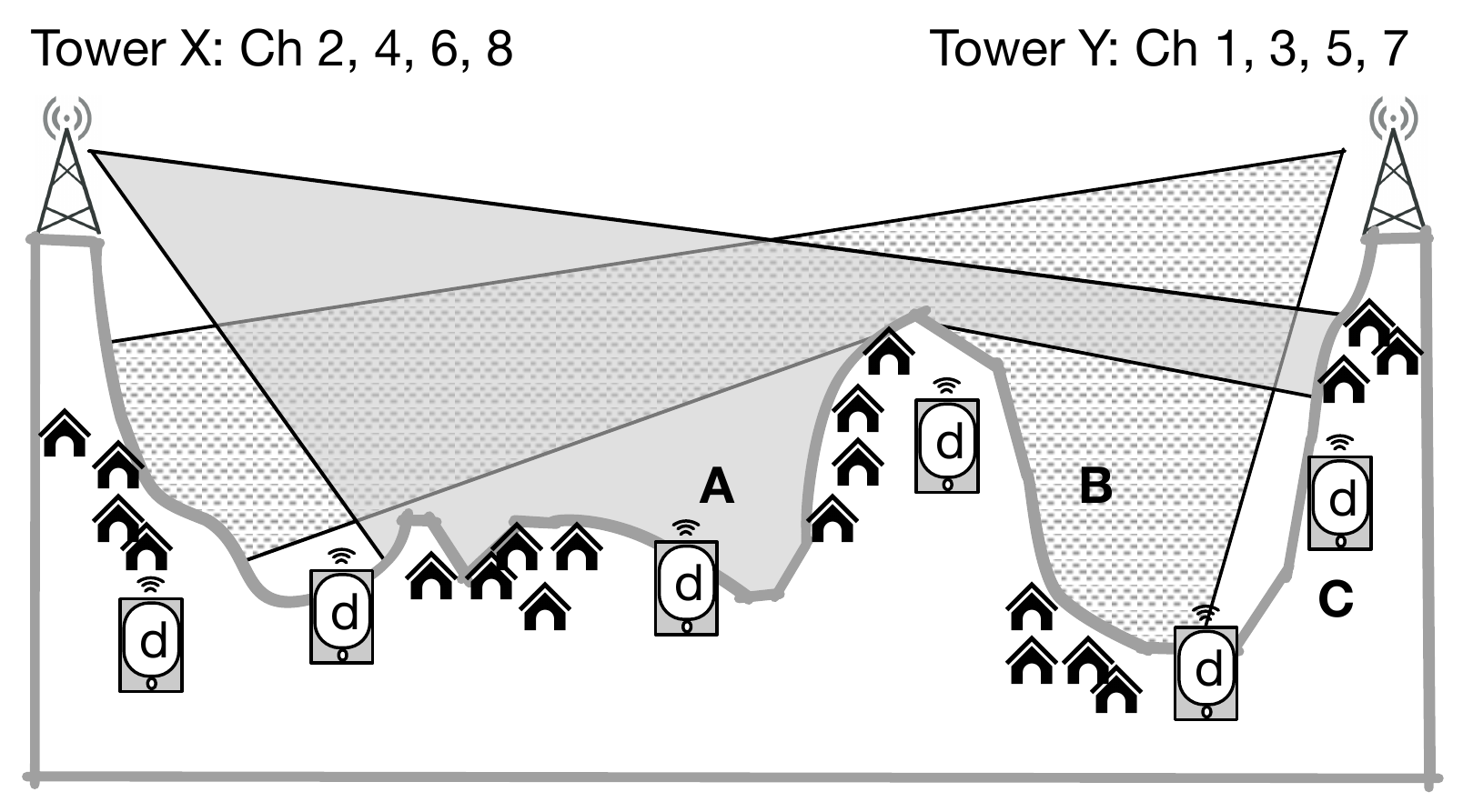}
\caption{Typical use-case for the detector (d) approach in White Spaces collection.}
\label{fig:detector-use-case}
\end{figure}

We are discussing a system implementing the detector approach, that offers a Geo-Location database for raw energy detection in the UHF band (but easily extensive to other bands), and beaconing for IEEE 802.11 networks. Once we have available raw data from the device, there is a collection process workflow that allows a better understanding of the spectrum from the community perspective. Fig.~\ref{fig:detector-use-case} shows the typical use-case for the detector (d) approach. With a mobile low-cost and low-weight device, a person can detect the presence of different signals coming from different (accounted or non-accounted) sources. In the figure, regions A and B can be clearly identified by the two different fingerprints collected with mobile devices. In this specific case, the separation of the two zones may be due to the planning of the operators. In region C, it may be harder to assess the presence of white spaces since Tower X and Y may overlap and produce false positives. Our proposed collection workflow is as follows. However, as suggested in the figure, it may exist in a deserted region of less interest for certain types of networks. 

In what follows, we describe the collection workflow.

\textbf{Planning.} For anticipating the regions to be observed and the configuration of the collector device. During this stage, there may be a mixture of interests coming from the community, the regulator, or the different operators. 

\textbf{Collection.} In this proposal, we are dealing with mobile collections. However, in particular sites of interest, such devices can also account for time series of spectrum dynamics to understand finer-grained spectrum occupancy.

\textbf{Uploading.} Once the spectrum is collected we can upload the raw files to \zebra which converts them into a compact JSON format with different types of (uploaded) traces. Once in the database, \zebra can provide a filtered and better-formatted set of data. This feature brings not only storage savings but also, a more understandable and manageable format for the accounted collections. 

\begin{figure}
\vspace{0.5cm}
\includegraphics[scale=0.45]{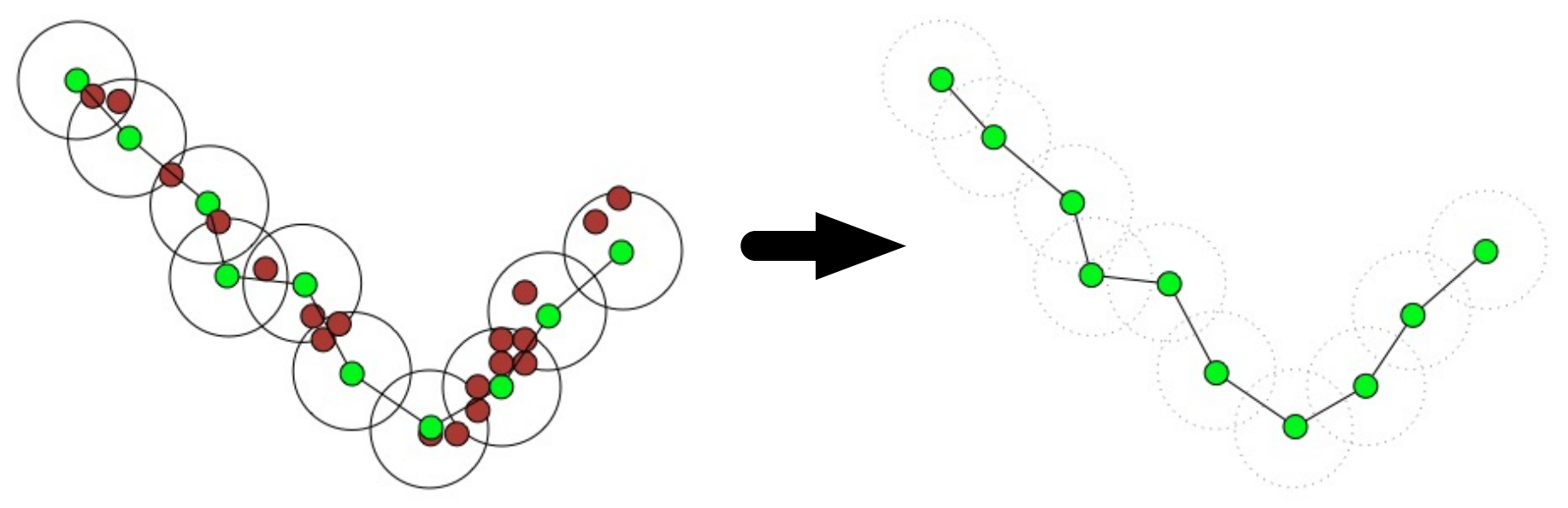}
\caption{Separation of collection points.}
\label{fig:cleaning-markers}
\end{figure}

\textbf{Processing.} \zebra can correct the bias produced during the collection process. An algorithm for evenly separating the collection points avoids the natural bias on convenient representations such as heat-maps. As shown in Fig.~\ref{fig:cleaning-markers}, \zebra provides a mechanism to separate collection points and avoid a biased collection. The left side of the figure shows a referential collection point at the center of every circle. The circumscription \textit{C} of the radius \textit{R} will define the area to be condensed\footnote{\footnotesize The most appropriate function should be applied to condense the spectrum data: max, min, average, etc.} into \textit{C}. The right side of the figure shows the result of the algorithm purging the uneven samples.

\textbf{Rezoning.} \zebra provides several representations so that the dataset can be further refined to leave the collection of interest. Depending on the user needs, a collection can be conveniently spaced (as in the previous step). \zebra also provides special functions for fine cutting the collection in a GUI with the scaled map of the zone.

\section{World-wide TVWS Collection}

%\begin{figure}
%\vspace{0.5cm}
%\includegraphics[scale=0.08]{img/world_map_hires.png}
%\caption{World map references of available reports}
%\label{fig:world-map}
%\end{figure}

In this section we present an extensive measurement campaign collected with different low-cost devices and from four different continents\footnote{\footnotesize The complete collection is freely available at http://wireless.ictp.it/tvws}. We make particular emphasis on developing regions, classified through the Internet Affordability Position provided by the Alliance for Affordable Internet (A4AI)\footnote{\footnotesize http://a4ai.org}. The ranking showed in the 4th column of Table~\ref{tab:summary-collections}, was obtained from the Affordability report as of 2014. 

%Fig.~\ref{fig:world-map} shows the different available regions from 4 different continents, collected and processed with \zebra for assessing the available white spaces. 
\begin{table}[htb]
\caption{Summary of collected journeys}
\centering
\begin{tabular}{|p{1.2cm}|p{2cm}|p{1.1cm}|p{1.2cm}|p{1cm}|}
%\begin{tabular}{|p{1.2cm}|p{2cm}|c|c|c|}
%{|>{\centering\arraybackslash}p{1.2cm}|>{\centering\arraybackslash}p{2cm}|>{\centering\arraybackslash}p{1.1cm}|>{\centering\arraybackslash}p{1.2cm}|>{\centering\arraybackslash}p{1cm}|}
\hline 
\textbf{Country}& \textbf{City} & \textbf{Total Distance (Km) }& \textbf{Internet Affordability Position } &\textbf{Avg. White Spaces}\\ 
\hline 
Costa Rica & Muelle, Santa Clara & 134.5 & 1 & 83\%\\
\hline 
Mauritius & Pereybere, Sottise, Valle des Pretres, Engrais Martial, Camp Caval,
Moka, Minissy, Saint Antoine & 93.4 & 7 & 48\%\\
\hline 
Ecuador & Puerto Aroya & 5.2 & 8 & 46\%\\
\hline 
Argentina & Ezeiza & 41.4 & 9 & 39\%\\
\hline 
Morocco & Chefchaouen Kasbah, Tahar, Douar Cheikh Driss, Ouled Sidi Chiekh,
Bou Touil & 44.7 & 12 & 46\%\tabularnewline
\hline 
Venezuela & Merida, Barquisimeto, El Vigia & 1000 & 37 & 86\%\\
\hline 
Mozam-bique & Boane, Sommerschield & 145.6 & 42 & 70\%\\
\hline \hline
Canada & Lunenburg & 0.1 & N/A & 86\%\tabularnewline
\hline 
Comoros & Anjouan, Grande Comore & 40.3 & N/A & 87\%\\
\hline 
Italy & Trieste & 0.1 & N/A & 74\%\tabularnewline
\hline 
Liberia & Central Monrovia, Kpegoa & 62.9 & N/A & 62\%\\
\hline 
\end{tabular} 
\label{tab:summary-collections}

\end{table}

%%\begin{table}
%\centering
%\caption{Frequency of Special Characters}
%\begin{tabular}{|c|c|l|} \hline
%Non-English or Math&Frequency&Comments\\ \hline
%\O & 1 in 1,000& For Swedish names\\ \hline
%$\pi$ & 1 in 5& Common in math\\ \hline
%\$ & 4 in 5 & Used in business\\ \hline
%$\Psi^2_1$ & 1 in 40,000& Unexplained usage\\
%\hline\end{tabular}
%\end{table}

In order to collect available TVWS, we used the so-called \textit{detector approach} \cite{brown}.  It consists on using different low-cost devices to scan the airwaves to detect the TV signals. The \textit{detector approach} is challenging since failures in detecting TV signals (caused by noise, other signals, etc.) or false positives due to atmospheric conditions pose particular limits to secondary users \cite{brown}. However, \zebra provides rapid means to see the amount of white spaces available for different threshold levels or to plot heat-maps of the available spectrum from uploaded mobile measurements. 

The assessment of White Spaces was done similar to the approach proposed in \cite{arcia}. Once we collected the spectrum measurement, they were uploaded into \zebra. Following the processing and rezoning steps recommended in Section~\ref{workflow}, we spaced the collection points considering a minimal variance of the consecutive separation of every possible neighbor pair. Then, we rezoned the markers to the well-known urban area. This rezoning was made with the intention of reporting the busiest occupation scenario in TV bands, thus making a stronger case for the use of TVWS in rural areas (always reporting a higher number of white spaces). The average white spaces are then calculated by fixing the occupation threshold of a journey to the value of the last channel reporting 100\% of occupation. This procedure is carried out by an observer scrolling the threshold on the \verb:zebra/place/occupation: plot. The ratio of White Spaces is then calculated as the number of channels with occupation lower than 20\% divided by the total possible channels on the UHF band. As shown in Table \ref{tab:summary-collections}, the amount of white spaces assessed with our approach accounts between 39\% and 86\%, in urban areas, of the A4AI report's countries. Surprisingly, well-known developed regions such as Canada or Italy show available white spaces of 74\% and 86\%, respectively, on the measured cities.

%\begin{figure}
%\vspace{0.5cm}
%\includegraphics[scale=0.18]{zebra-image.png}
%\caption{White Spaces report on Zebra-RFO}
%\label{fig:zebra-img}
%\end{figure}

%Fig.~\ref{fig:zebra-img} shows the operating prototype of 

\section{Conclusions}

In this paper, we have argued the need for open and regionalised repositories for managing the spectrum occupancy. We have broadly discussed an architecture supporting the different needs for a two-tier approach to the governance of the spectrum. Furthermore, we present a system allowing the storage of long-term spectrum dynamics produced by low-cost devices. \zebra offers editing capabilities on geo-tagged data that allows the observation of an area of interest for network deployment. Successful experiences suggest that the use of low-cost devices and useful visualisation are of help, providing regulators and crowds with easy-to-digest information, and lowering the costs of spectrum collection by, delegating the task of finding the use-cases to the people. 

We are currently working towards the use of intelligent techniques such as machine learning or fuzzy logic to assess white spaces automatically and to provide prospective use of the spectrum. We will also look at the potential combining \zebra data with TV channel availability results from a geo-location spectrum database to assess the best TV channels to use for secondary users. Some statistical inference may be needed if \zebra spectrum data is not available at the exact GPS location being checked. Finally, we consider this initiative as a fundamental step to bringing the next billion people on-line. 

\section*{Acknowledgements}
The research leading to these results has received funding from the European Union\textquotesingle s (EU) Horizon 2020 research and innovation programme under grant agreement No. 644663. Action full title: architectuRe for an Internet For Everybody, Action Acronym: RIFE.

\bibliography{biblio-openrep}
\end{document}